\documentclass{article}
\usepackage{amsmath}
\usepackage{amsfonts}
\usepackage{amssymb}
\usepackage{epsfig}
\usepackage{wrapfig}
\usepackage{siunitx}
\usepackage{authblk}

\usepackage{color}

\newcommand{\ie}{{i.e.}~}

\newcommand{\Tw}{\text{Tw}}
\newcommand{\Wr}{\text{Wr}}
\newcommand{\Lk}{\text{Lk}}
\newcommand{\beq}{\begin{equation}}
\newcommand{\eeq}{\end{equation}}
\newcommand{\ben}{\begin{eqnarray}}
\newcommand{\een}{\end{eqnarray}}

\begin{document}

\title{Modeling Bacterial DNA: Simulation of Self-avoiding Supercoiled Worm-Like Chains Including Structural Transitions of the Helix}

\author{Thibaut Lepage and Ivan Junier\thanks{Correspondence should be sent at: \texttt{ivan.junier@univ-grenoble-alpes.fr}}}
\affil{CNRS, TIMC-IMAG, F-38000 Grenoble, France\\Univ. Grenoble Alpes, TIMC-IMAG, F-38000 Grenoble, France}

\date{}

\maketitle

\section*{Abstract}
Under supercoiling constraints, naked DNA, such as a large part of bacterial DNA, folds into braided structures called plectonemes.
The double-helix can also undergo local structural transitions, leading to the formation of denaturation bubbles and other alternative structures.
Various polymer models have been developed to capture these properties, with Monte-Carlo (MC) approaches dedicated to the inference of thermodynamic properties.
In this chapter, we explain how to perform such Monte-Carlo simulations, following two objectives.
On one hand, we present the self-avoiding supercoiled Worm-Like Chain (ssWLC) model, which is known to capture the folding properties of supercoiled DNA, and provide a detailed explanation of a standard MC simulation method.
On the other hand, we explain how to extend this ssWLC model to include structural transitions of the helix.

\section*{Key Words}
Monte-Carlo Methods, DNA Supercoiling, Worm-Like Chain, Plectonemes, DNA Denaturation, Structural Transitions, Multi-Scale Simulations

\section{Introduction}

In bacteria, chromosomes are partly structured by DNA supercoiling.
Compared to its natural helicity, DNA is indeed often found in an underwound form {\it in vivo}, as a result of a yet-to-be-understood balance between transcription, replication and the action of topoisomerases and nucleoid associated proteins~\cite{Thanbichler:2005hc,Blot:2006gh,Lagomarsino:2015kg}.
On one hand, DNA supercoiling leads to the formation of plectonemes.
This can be explicitly shown using Worm-Like Chain (WLC) models of DNA that include supercoiling constraints and self-avoidance properties reflecting the impenetrable character of the DNA molecule~\cite{Vologodskii1994,Vologodskii1997,Klenin:1998gq}. Specifically, supercoiling constraints make molecules buckle so that they absorb, under the form of writhe, some of the excess or depletion of twist, while the short-range repulsion (self-avoidance) results in an effective entropic repulsive force that determines the radius of plectonemes~\cite{Marko1995}.
Self-avoiding supercoiled WLC (ssWLC) models are thus able to capture the folding properties of supercoiled DNA on the scale of several kilo base pairs (kbp), or tens of kbps~\cite{Lepage2015}, both for overwound (positive supercoiling) and underwound (negative supercoiling) DNA at low tension forces. On the other hand, negative supercoiling can induce structural transitions towards DNA forms different from the canonical B-DNA. Among a large number of possible alternative structures~\cite{Mirkin:2008uj}, supercoiled DNA can locally form denaturation bubbles~\cite{Benham:1992tg,Strick:1998ww} or adopt left-handed DNA forms such as Z-DNA~\cite{Oberstrass:2012jc} or the so-called L-DNA~\cite{Sheinin2011,Vlijm2015}.

Since most bacterial chromosomes are negatively supercoiled, models dedicated to biological applications are destined to capture the balance between super-structuring (plectonemes) and local structural changes of the DNA-helix. Several recent approaches have thus been proposed to tackle this multi-scale problem. These include phenomenological models of the co-existence of the plectonemic and denatured states~\cite{Marko:2007cf,Oberstrass:2012jc,Meng2014} as well as polymer models at the resolution of a single base ~\cite{Matek:2015gb} (see ~\cite{Manghi2016} for a recent review of single-base based models).

Here, we will explain how to design a discrete version of the ssWLC model in order to simulate the competition between plectoneme formation and structural transitions of the DNA helix, resulting in a 10-to-20-bps resolution model that can be used to simulate several kbps long molecules under negative supercoiling~[Lepage \& Junier, in prep]. To this end, we first aim at recalling the definition of the discrete version of the ssWLC and at providing a detailed description of the methods used to simulate its equilibrium folding properties.

\section{Methods}

Methods are organised as follows. First, we recall the definition of the ssWLC and explain the discretisation procedure to simulate it. We next explain how to parametrize the fundamental units of the model in order to solve the problem of the conservation of the linking number, which is at the root of the supercoiling constraints. We then recall principles of thermodynamics-oriented Monte-Carlo methods and discuss the problems of the detection of collisions and chain crossing, which are the most time-consuming steps of the simulations. Finally, we explain how to include structural transitions.

\subsection{The discrete self-avoiding supercoiled WLC model}

\begin{figure}[t]
	\begin{center}
    \includegraphics[width=0.8\textwidth]{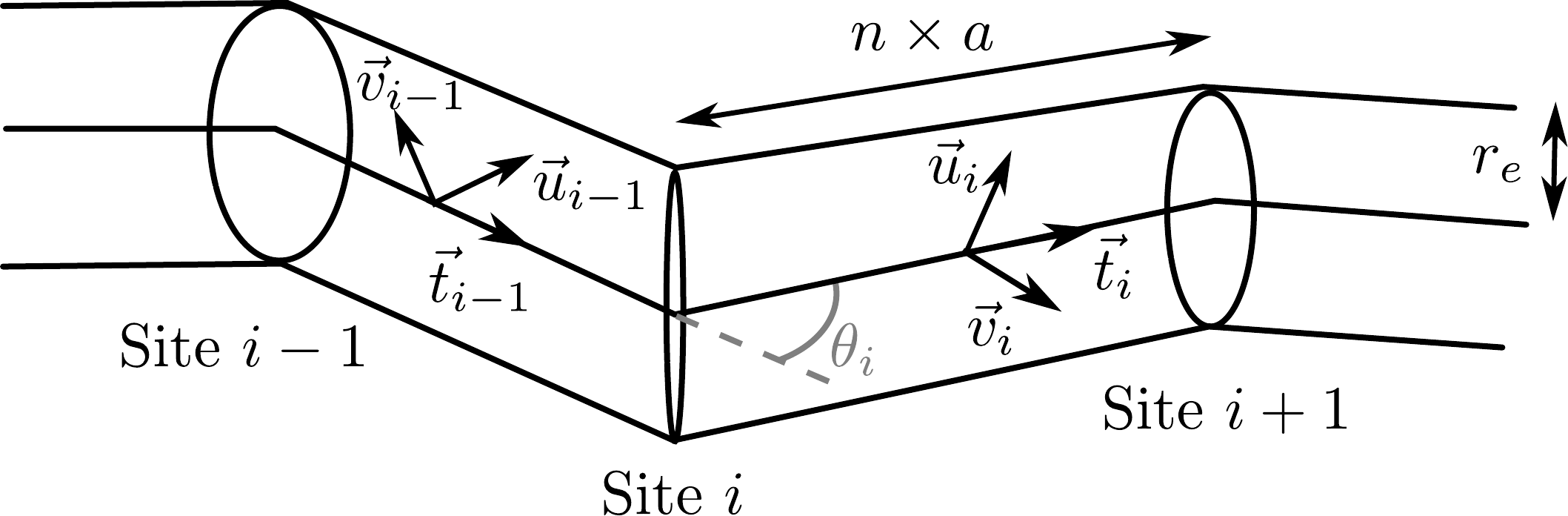}
    \caption{A discrete model of the ssWLC. Here, two complete cylinders surrounding a site $i$ are represented, along with their local frames $(\vec{t}, \vec{u}, \vec{v})$ used to compute the bending angle $\theta_i$ and the twist angle $\phi_i$ (not indicated, see text).}
	\label{fig:model}
	\end{center}
\end{figure}

A WLC model provides a continuous description of a polymer chain. At the microscopic level, the simplest model is defined by a single bending modulus characterising the cost for the chain to locally bend. The chain then has a certain persistence length ($\ell_p$), below which it keeps memory of its orientation. Typically, $\ell_p\approx\SI{50}{nm}$ for B-DNA at physiological salt concentration.

To investigate the spatial properties of a WLC, one generally resorts to a discretisation procedure, which consists in dividing the chain into a succession of $N-1$ identical segments articulated by $N$ sites (Figure~\ref{fig:model}), counting the two sites located at the extremities of the chain -- in the case of a circular molecule (e.g. a plasmid), there are as many segments as sites, and the first site joins the last and first segments together. For DNA molecules in physiological conditions of salt, an additional classical simplification consists in considering a hard-core description of the electrostatic repulsion of the negatively charged DNA backbone, such that segments of the discrete WLC are in fact impenetrable cylinders characterized by a radius $r_e$ (Figure 1), with $r_e$ depending on the salt concentration~\cite{Rybenkov1997} (for instance $r_e=\SI{2}{nm}$ for [NaCl]=\SI{100}{mM}). Altogether, this framework defines the discrete self-avoiding WLC.

In this context, each site $i$ is associated with a "discrete" bending modulus ($K_{d}$), which constrains the amplitude of the bending angle ($\theta_i$, see next section for an operational definition) between the tangent vectors of the cylinders $i-1$ and $i$ (Figure~\ref{fig:model}), and whose value is adjusted depending on the level of discretisation -- note here that "the cylinder $i$" corresponds to the cylinder located between the sites $i$ and $i+1$ (for simplicity, we drop the reference to the sites and cylinders in the indexing). Specifically, denoting $k_B$ the Boltzmann constant, $T$ the temperature ($T=\SI{310}{K}$ in physiological conditions), $n$ the number of bps per cylinder and $a$ the average distance between any two consecutive bps ($a=\SI{0.34}{nm}$ for B-DNA) such that $na$ is the length of a cylinder, the associated persistence length reads $\ell_p=na\frac{K_{d}}{k_BT}$. For a given value of $\ell_p$, then, the smaller the discretization is, the larger the value of $K_{d}$. Note, here, that it is recommended to work with cylinders small enough to avoid discretisation artefacts, a typical choice being 5 cylinders per $\ell_p$~\cite{Vologodskii1997} such that $n=30\text{bps}$ for B-DNA (see Note~\ref{coarsegraining}).

In the case of a ssWLC model, a torsional modulus $C$ must additionally be considered to account for the cost associated with the twist deformation of the chain ($C$ is typically on the order of $\SI{100}{nm}$ for B-DNA~\cite{Strick1999,Bryant2003}). In this context, an average twist angle ($\phi_i$, see next section for an operational definition) is associated to each site $i$ of the chain~\cite{Wu:1988vs,Klenin:1998gq}. $\phi_i$ is equal to the average twist angle between the $n$ base pairs located closest to the site $i$. Then, just as $\theta_i$ is constrained by the bending modulus, $\phi_i$ is constrained by the torsional modulus (see Eq.~\ref{eq:E}).

Altogether, every conformation $\mathcal{C}$ of the chain has an intrinsic thermodynamic weight that depends on the associated bending and torsional properties. This results in a conformational energy, $E(\mathcal{C})$, which, in the presence of a stretching force $f$ (as in the case of single-molecule experiments), reads (Note~\ref{global}):
\beq
E(\mathcal{C})=\frac{k_BT}{2}\sum_{i=1}^N\left[\frac{\ell_p}{na}\theta_i^2+\frac{nC}{a}(\phi_i-\phi_0)^2\right]-fz,
\label{eq:E}
\eeq
with $z$ the extension of the chain along the axis of the force, and $\phi_0$ the average twist angle at rest, that is, the unconstrained helicity of DNA ($\phi_0=0.6$ for B-DNA). Note also that the bending angles for the extreme sites $i=1$ and $i=N$ are defined with respect to the axis of the force.

Finally, to simulate the folding properties of such a ssWLC, topological constraints immanent to the double-stranded nature of DNA must be accounted for. Namely, for both circular DNA molecules and linear molecules whose ends cannot rotate, the linking number $\Lk(=\Tw+\Wr)$ is an invariant quantity, meaning that the sum of the twist $\Tw=(2\pi)^{-1}\sum_{i=1}^N\phi_i$ (the number of helices) plus the writhe $\Wr$ (the number of loops made by the axis of the molecule around itself, see Note~\ref{lk}) remains always the same, unless the molecule is cut by, e.g., an enzyme. In other words, should DNA be unwound (or overwound), the net change of the linking number will be distributed between the twist and the writhe. From a simulation viewpoint, this imposes strong  constraints on the definition of the twist angle for the discrete ssWLC, as we now explain.

\subsection{Bending and twist angles}

Denoting $\vec{t}_i$ the tangent vector of the cylinder $i$, the bending angle $\theta_i$ between $i-1$ and $i$ is unambiguously given by (Figure~\ref{fig:model}):
\beq
\cos\theta_i=\vec{t}_{i-1}\cdot\vec{t}_i
\label{eq:bending}
\eeq

Due to the fact that the ssWLC does not include any explicit representation of the DNA helix, the definition of the twist angle has been more ambiguous. Methods using Euler angles between local frames associated to the cylinders were first proposed, and shown to provide an excellent procedure to have a linking number that fluctuates around a fixed value~\cite{Wu:1988vs}. More recent methods~\cite{Bergou:2008va,Carrivain2014,Lepage2015} based either on an explicit representation of the double helix~\cite{Lepage2015}, or on a definition of the twist angle coming from the parametrisation of the deformation of rigid bodies~\cite{Bergou:2008va}, allow to conserve the linking number exactly during the simulations.
In particular, using the ``parallel transport'' approach developed in~\cite{Bergou:2008va}, it is possible to define a twist angle unambiguously such that after each block rotation of the cylinders (see below), the linking number remains constant. To this end, for every site $i$, define at the initial time of the simulation a vector orthogonal to $\vec{t}_i$, here called $\vec{u}_i$ (Figure~\ref{fig:model}). Next, just as for the $\vec{t}_i$'s, continuously update these $\vec{u}_i$'s by applying the rotation matrix corresponding to the deformation of the chain (see below). In this context, using $\vec{v}_i=\vec{t}_i\times\vec{u}_i$, $\phi_i$ is given by~\cite{Carrivain2014}:
\begin{gather}
\cos\phi_i=\frac{\vec{u}_{i-1}.\vec{u}_i+\vec{v}_{i-1}.\vec{v}_i}{1+\vec{t}_{i-1}.\vec{t}_i}\\
\sin\phi_i=\frac{\vec{v}_{i-1}.\vec{u}_i-\vec{u}_{i-1}.\vec{v}_i}{1+\vec{t}_{i-1}.\vec{t}_i}.
\label{eq:twist}
\end{gather}

The exact conservation of the linking number can then be verified explicitly by computing the twist and the writhe (see Note~\ref{lk}). In practice, the writhe involves a sum over all pairs of cylinders and, hence, should be computed only once in a while.

\subsection{Monte-Carlo method: elementary moves, transition probabilities and ergodicity properties}

\begin{figure}[t]
	\begin{center}
	\includegraphics[width=0.4\textwidth]{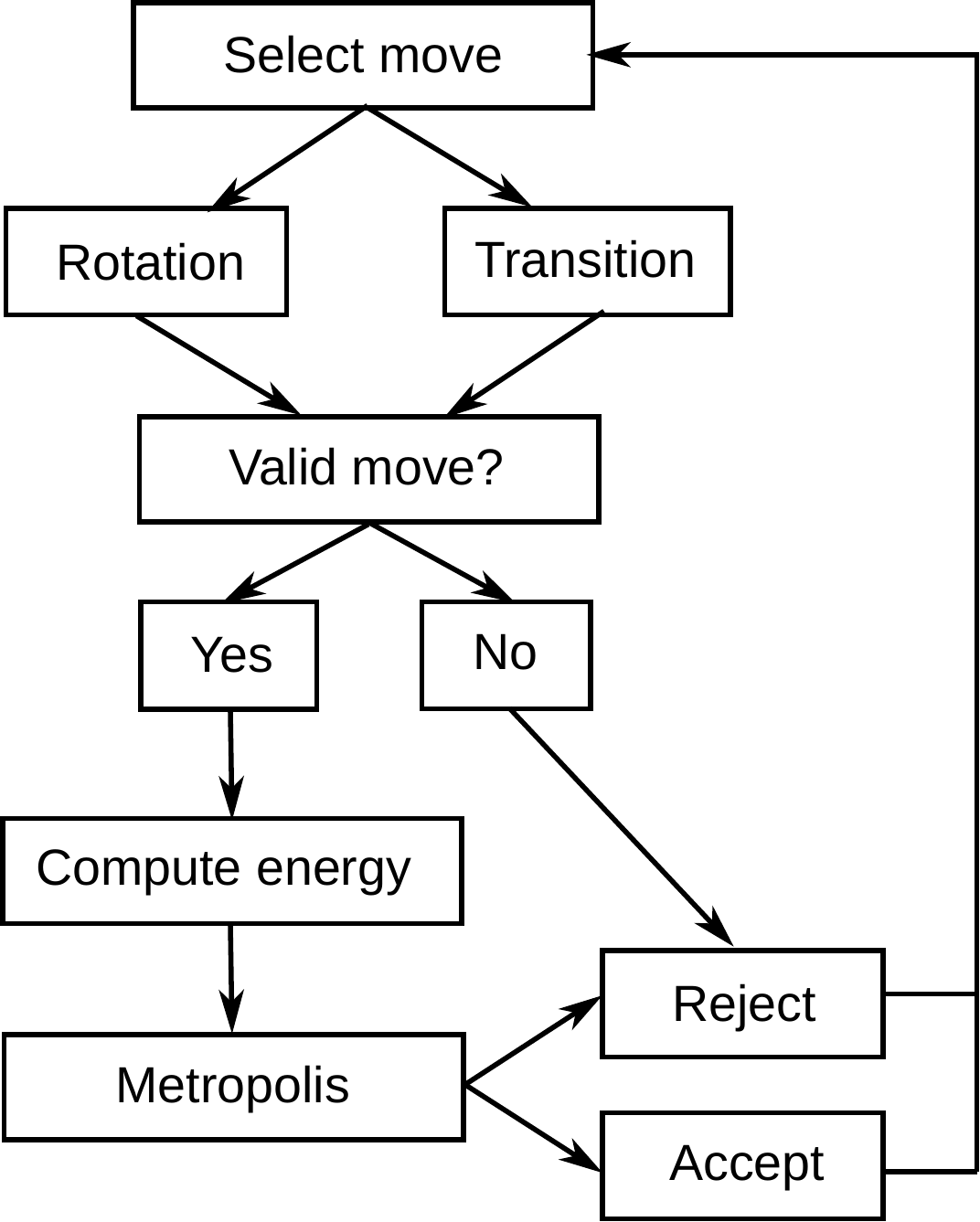}
    \caption{Summary of the Monte-Carlo Metropolis algorithm.
    Note that the ssWLC requires only rotations, so that there is no need to select the type of elementary move to try in this case.
    The second type of move is used to take structural transitions into account.
    The elementary moves are described in Sections~\ref{R} and~\ref{ST}, and the validity check in Section~\ref{DCC}.}
	\label{fig:MC}
	\end{center}
\end{figure}

Given the energy of the conformations (Eq.~\ref{eq:E}), the hard-core repulsion of the chain (self-avoidance) and the constraint of the conservation of $\Lk$, the equilibrium sampling of the conformations is usually performed using a Monte-Carlo (MC) method. This consists in generating a large number of successive conformations (Figure~\ref{fig:MC}) such that the occurrence of conformations with energy $E$ eventually becomes proportional to $\mathcal{N}(E)\exp(-E/k_BT)$ (Boltzmann's law), where $\mathcal{N}(E)$ is the total number of conformations with energy $E$. In practice, it is achieved by constructing each conformation from the previous one via trials of elementary moves, starting from an initial random conformation~(see Note~\ref{IC}). To this end, the transformation of a conformation $\mathcal{C}$ into another conformation $\mathcal{C}'$ is accepted with a certain transition probability $W(\mathcal{C}\to\mathcal{C}')$ if no collision or crossing is generated during the applied elementary move. Different forms of transition probabilities can be used, the only constraint being that these must verify the detailed balance condition,
$W(\mathcal{C}\to\mathcal{C}')/W(\mathcal{C}'\to\mathcal{C})=e^{-\frac{E(\mathcal{C}')-E(\mathcal{C})}{k_BT}}$, which ensures that at large enough time the chain will visit conformations according to their Boltzmann weight. In this regard, a classical choice is the Metropolis-Hastings transition rate~\cite{Metropolis1953,Newman1999}, such that:
\beq
W(\mathcal{C}\to\mathcal{C}')=\max\left\{1,\exp\left[-\left(E(\mathcal{C}')-E(\mathcal{C})\right)/k_BT\right]\right\}.
\label{eq:detbal}
\eeq
Note that in this case any elementary move decreasing the energy is accepted.

A sufficiently high number of successive elementary moves must then be generated in order to get enough uncorrelated conformations such that the sampling of the conformations is representative of thermodynamic equilibrium.
It is also important to check that the conformations visited during the simulation do not correspond to a metastable state only (the so-called ergodicity problem), that is, that the system is not trapped in a subset of conformations whose free energy is on the same order of magnitude as that of another ``unreached'' subset.
To this end, it is often convenient to use an ``annealing procedure'', which consists in starting with a value of some parameter (usually the temperature $T$) such that the system can quickly reach equilibrium, and then varying this value progressively until the working value is reached.
The supercoiling level $\sigma$ is also a parameter well fitted for this method: starting with a torsionally relaxed molecule, one can perform simulations at various constant values of  $\sigma$  by continuously increasing (or decreasing) its value, using the last conformation of each simulation as the initial condition of the next one.
Then, with good confidence, equilibrium is reached if the statistical properties of the resulting conformations at the working value do not depend on the speed at which the annealing has been realised. On the opposite case, one needs to resort to other types of elementary move~\cite{Liu:2008hx}, or other techniques of simulations to prevent the system from being trapped in specific states (see e.g.~\cite{2008RPPh...71l6601L}).

\subsection{Rotations and associated change of bending and torsional properties}
\label{R}

\begin{figure}[t]
	\begin{center}
    \includegraphics[width=0.7\textwidth]{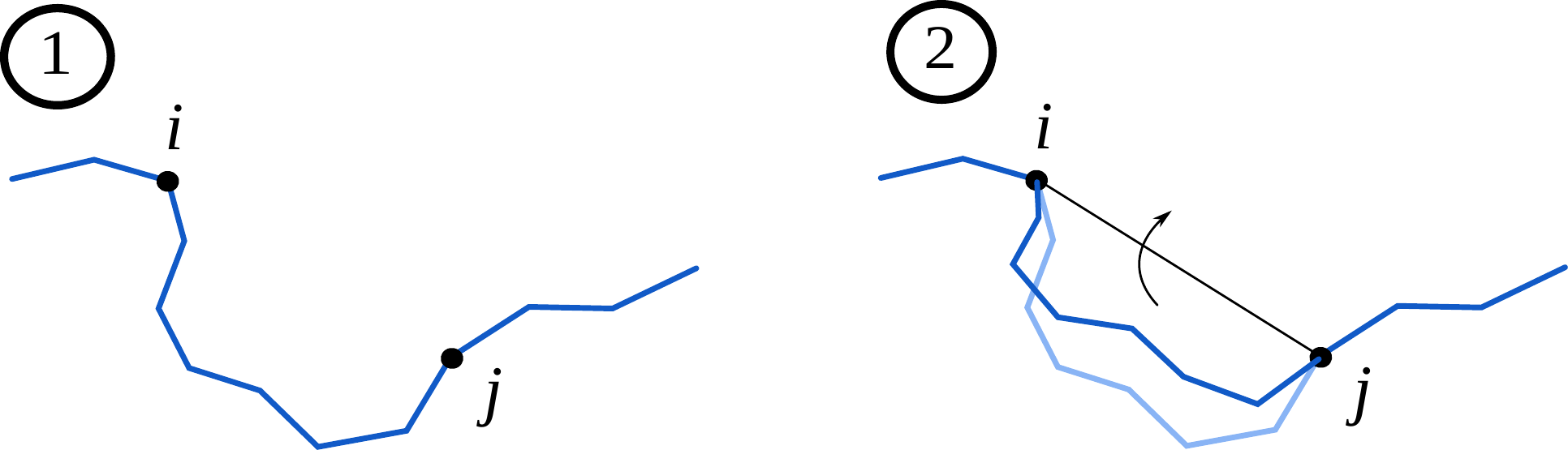}
    \caption{Rotation: pick a random block $[i,j]$ of cylinders (1) and rotate it around its axis according to a random angle (2).}
    \label{fig_rotation}
	\end{center}
\end{figure}

Most commonly in the MC simulation of a WLC model, an elementary move consists in randomly rotating a block of contiguous cylinders, also called a crankshaft move (Figure~\ref{fig_rotation}).
To this end, first define once for the entire simulation a maximum number ($M$) of cylinders allowed to rotate simultaneously and a maximum angle of rotation ($\alpha$) (see Note~\ref{acceptance}). A rotation then consists in (i) picking a site $i$ at random, (ii) picking a number $m<M$ at random, (iii) choosing a direction $s\in\{-1,1\}$, (iv) defining the axis of the rotation as the straight line connecting the two sites $i$ and $j=i+s\times m+1$ (modulo the number of sites), (v) choosing an angle of rotation randomly in $[-\alpha,\alpha]$, and (vi) applying the corresponding rotation matrix to the vectors $\vec{t}$ and $\vec{u}$ associated to each site of the block $[i,j]$ (Figure~\ref{fig_rotation}).

After having applied the rotation, test whether a collision or a crossing occurred (explained in Section~\ref{DCC}). If this is the case, the rotation is rejected (see Note~\ref{ptrs}). In the opposite case, use Eq.~\ref{eq:bending} and Eq.~\ref{eq:twist} to compute the new bending and twist angles at sites $i$ and $j$ (borders of the block). Then, compute the corresponding energies (see Eq.~\ref{eq:E}) to obtain the total energy variation generated by the rotation. Finally, accept the rotation according to the transition probability $W$ (see Eq.~\ref{eq:detbal}). 

For linear molecules, when the site $i$, the direction $s$ and the size $m$ of the block are such that the extreme site $j$ falls outside the chain ($j\le0$ or $j\geq N+1$), the crankshaft rotation becomes ill-defined. In this case, apply a rotation to the block extending from $i$ to the end of the molecule (site $1$ or site $N$, depending on the direction $s$), around a random axis and according to an angle chosen at random in $[-\alpha,\alpha]$. Note that this type of elementary move is necessary to displace the end points of the linear chain; in such case, the extension ($z$) and, hence, the corresponding stretching energy ($-fz$) must also be updated. In the case of a linear molecule for which the linking number needs to be conserved, extra precaution must also be taken. In particular, it is necessary to define two walls bound to each end of the molecule and perpendicular to the stretching force~\cite{Vologodskii1997,Lepage2015}, and to prevent any cylinder from trespassing them (see Note~\ref{walls}). These walls can be viewed as a simple modelling of the fixed surface and the magnetic bead used in single-molecule experiments.

\subsection{Detection of Collisions and Crossings}
\label{DCC}

To detect both collisions and crossings, first build a mesh of the volume inside which the molecule is embedded, using a width of cells ($w_\text{cell}$) larger than $na+2r_e$, so that collisions may only occur between cylinders whose center are in the same cell or in nearest neighbours (cells sharing a face, an edge or a vertex). The collision detection for a cylinder $i$ that has just moved (as a consequence of the trial of an elementary move) then consists in testing whether $i$ collides with one of the cylinders not belonging to the block but located in one of the $27$ ($=3^3$) cells in the vicinity of $i$. Note, here, that cylinders at a distance too close to $i$ along the chain (\ie separated from $i$ by less than $2r_e$ when measured along the axis of the chain) must be ignored since, by construction, they always overlap with $i$ (see Note~\ref{coarsegraining}).
Then, to test whether the new position $\vec{r}_i$ of $i$ leads to a collision with a cylinder $j$ at position $\vec{r}_j$, test whether $(\vec{r}_j-\vec{r}_i).\vec{n}>2r_e$, where $\vec{n}=\vec{t}_i\times\vec{t}_j$ is the normal to the plane $(\vec{t}_i,\vec{t}_j)$. If the inequality holds, then there is no collision. In the opposite case, check first the distances between the four possible pairs of cylinder ends. If one of these distances is smaller than $2r_e$, then there is a collision. In the opposite case, there is still a possibility of collision in the middle of both cylinders.
The intersection of the two straight lines $(\text{Site } i, \vec{t}_i)$ and $(\text{Site } j, \vec{t}_j)$ is located at the abscissa $u$ along $(\text{Site } i, \vec{t}_i)$ and $v$ along $(\text{Site } j, \vec{t}_j)$, where:
\begin{gather}
u=(\vec{r}_j-\vec{r}_i)\frac{\vec{t}_i-(\vec{t}_i.\vec{t}_j)\vec{t}_j}{1-(\vec{t}_i.\vec{t}_j)^2}\\
v=-(\vec{r}_j-\vec{r}_i)\frac{\vec{t}_j-(\vec{t}_i.\vec{t}_j)\vec{t}_i}{1-(\vec{t}_i.\vec{t}_j)^2}.
\end{gather}
Then, the cylinders collide if and only if $0<u<na$ and $0<v<na$.

\begin{figure}[t]
	\begin{center}
    \includegraphics[width=\textwidth]{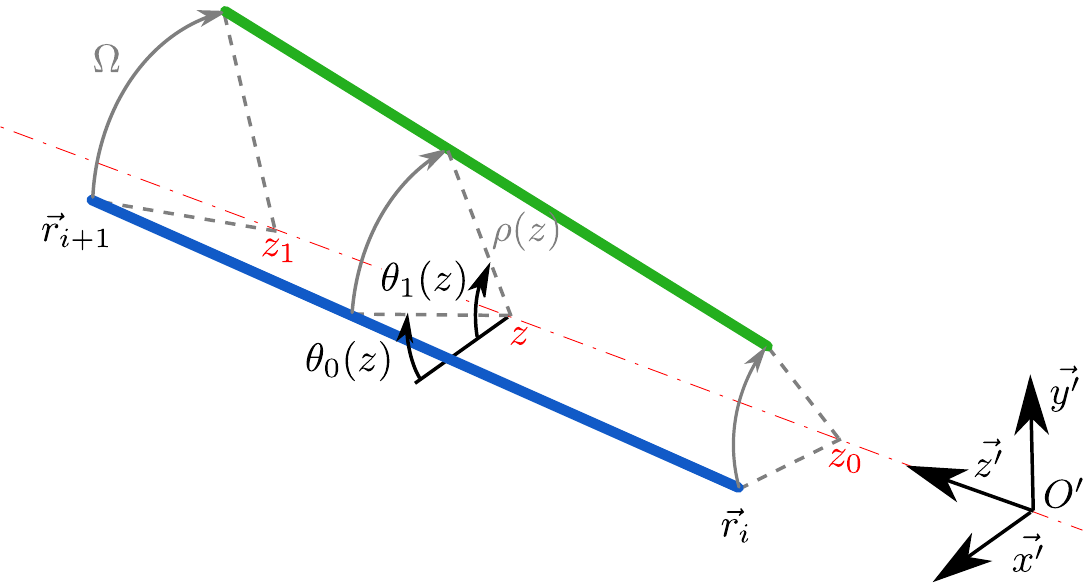}
    \caption{Schematic representation of the rotation of the support segment of a cylinder (thick lines) from the blue position to the green one.
    Each point of the segment moves with constant coordinates $z$ and $\rho(z)$ in the frame $(O',\vec{x'},\vec{y'},\vec{z'})$.
    The coordinate $\theta$ goes from $\theta_0(z)$ to $\theta_1(z)=\theta_0(z)+\Omega$, where $\Omega$ is the angle of the rotation. The red dashed line indicates the axis of rotation.}
    \label{fig_cone}
	\end{center}
\end{figure}

Next, in order to detect crossing events during the rotation of a cylinder $i$, first build a list of all the cells that may contain a cylinder crossed by $i$.
This list corresponds to the cells whose center lies at a distance smaller than $a+\frac{\sqrt{3}}{2}w_\text{cell}$ from the arc formed by the rotation of the center of $i$ (within these cells, just as in the case of collisions, only the non-moving cylinders have to be processed).
Next, define a frame $(O',\vec{x'},\vec{y'},\vec{z'})$ associated with the rotation of $i$~(Figure~\ref{fig_cone}), i.e. set the origin $O'$ anywhere on the axis of the rotation and align $\vec{z'}$ with this axis.
Then define $\vec{x'}$ and $\vec{y'}$ arbitrarily in order to complete an orthonormal base (for example, in Figure~\ref{fig_cone}, $\vec{x'}$ points toward $\vec{r}_i$).
In that frame, the surface swept by the support segment of the cylinder $i$ during the rotation is bound by $z_0$ and $z_1$, the coordinates of its ends along $\vec{z'}$~(Figure~\ref{fig_cone}).
For any given $z$ in this interval, the cylindrical coordinates of this surface are also easy to determine: $\rho(z)$ is constant and $\theta$ lies between $\theta_0(z)$ and $\theta_1(z)$, the initial and final angular coordinates at $z$.
Thus, the coordinates $(\rho_I,\theta_I,z_I)$ of a crossing point (intersection between the support segment of a cylinder $j$ and the surface swept by that of $i$) must verify:
\begin{gather}
\label{eq_zi}
z_0<z_I<z_1\\
\label{eq_rho}
\rho_I=\rho(z)\\
\label{eq_theta}
	\begin{cases}
	\theta_0(z_I)<\theta<\theta_1(z_I)&\text{if $\theta_0(z_I)<\theta_1(z_I)$}\\
	\theta<\theta_0(z_I)\text{ or }\theta>\theta_1(z_I)&\text{otherwise}
	\end{cases}
\end{gather}
In addition, this intersection must belong to the support segment of $j$, i.e. lie on its support line:
\begin{gather}
\label{eq_rcost}
\rho_I\cos\theta_I=x_j+\frac{t_{jx}}{t_{jz}}(z_I-z_j)\\
\label{eq_rsint}
\rho_I\sin\theta_I=y_j+\frac{t_{jy}}{t_{jz}}(z_I-z_j)
\end{gather}
and be bound by the ends of $j$:
\begin{gather}
\label{eq_zj}
	\begin{cases}
	z_j<z_I<z_{j+1}&\text{if $z_j<z_{j+1}$}\\
	z_{j+1}<z_I<z_j&\text{otherwise},
	\end{cases}
\end{gather}

Equations (\ref{eq_rho}), (\ref{eq_rcost}) and (\ref{eq_rsint}) lead to a quadratic equation for $z_I$, the solutions of which (if any) yield $\rho_I$ and $\theta_I$. Then, if exactly one solution verifies the inequalities~(\ref{eq_zi}), (\ref{eq_theta}) and~(\ref{eq_zj}), a crossing occurred and the trial must be rejected (see Note~\ref{cross_sense}).

\subsection{Including Structural Transitions}
\label{ST}

Structural transitions of the DNA-helix can be included and simulated in order to capture the multi-scale properties of negatively supercoiled molecules~[Lepage \& Junier, in prep]. To this end, one first needs to modify the form of the conformational energy (Eq.~\ref{eq:E}) to account for the possibility that the sites can now be associated to different DNA-forms (called states hereafter). This is done by adding a new variable, $s_i$, reflecting the state of the  site $i$~\cite{Manghi:2009el}, such that for instance $s_i=\text{B, D or Z}$ if one considers B-DNA (B), denaturation bubbles (D) and Z-DNA (Z). Importantly, compared to the previous single-state case, every state $s$ has its own mechanical properties: $\ell_s$ (persistence length), $C_s$ (torsional module), $\phi_{0,s}$ (average twist angle at rest) and $a_s$ (distance separating consecutive bps) -- we consider, for simplicity, a single electrostatic radius $r_e$ independent of states. In addition, the non-B states are characterized by a free energy formation per bp, denoted $\gamma_s$, reflecting the deformation of the base-pairing and stacking of the base pairs, which is on the order of $k_BT$ per bp~\cite{SantaLucia:1998uz}. Finally, one must consider a domain wall penalty, $J$, corresponding to the energy between any two sites having different states~\cite{Manghi:2009el,Oberstrass:2013du}. This term reflects the energy cost to go from one DNA-form to another and constrains the alternative forms to produce as few domains as possible (see~\cite{Manghi:2009el,Oberstrass:2013du} for further details). Altogether, the new energy ($E'(\mathcal{C})$) of a conformation $\mathcal{C}$ with multiple possible forms along the chain reads:
\beq
E'(\mathcal{C})=\frac{k_BT}{2}\sum_{i=1}^N\left[n\gamma_{s_i}+\frac{\ell_{s_i}}{na_{s_i}}\theta_i^2+\frac{nC_{s_i}}{a_{s_i}}(\phi_i-\phi_{0,s_i})^2\right]+J\sum_{i=1}^{N-1}\delta_{s_i,s_{i+1}}-fz.
\label{eq:Es}
\eeq
where we considered $\gamma_B=0$ (reference form) and where $\delta_{s_i,s_{i+1}}=1$ if $s_i=s_{i+1}$, $0$ otherwise.

Values of $J$, $C_s$ and $\ell_s$ have been estimated using single-molecule experiments~\cite{Sheinin2011,Oberstrass:2012jc,Oberstrass:2013du,Vlijm2015}. In addition, $\phi_{0,D}=0$ (by definition), $\phi_{0,Z}=0.52$ and $a_Z=\SI{0.37}{nm}$~\cite{Rich:1984ec} (from crystallographic measurements), while $a_D=\SI{0.54}{nm}$  has been previously used~\cite{Sheinin2011}. In this regard, because $a_s$ differs from state to state, one must now consider different sizes of the cylinders.
In practice, we use $n(a_i+a_{i+1})/2$ for the length of the cylinder defined by the sites $i$ and $i+1$. As a consequence, each time the state of a site $i$ is updated, the length of the two surrounding cylinders has to be updated, which is done in the following way.

\begin{figure}[t]
	\begin{center}
    \includegraphics[width=0.7\textwidth]{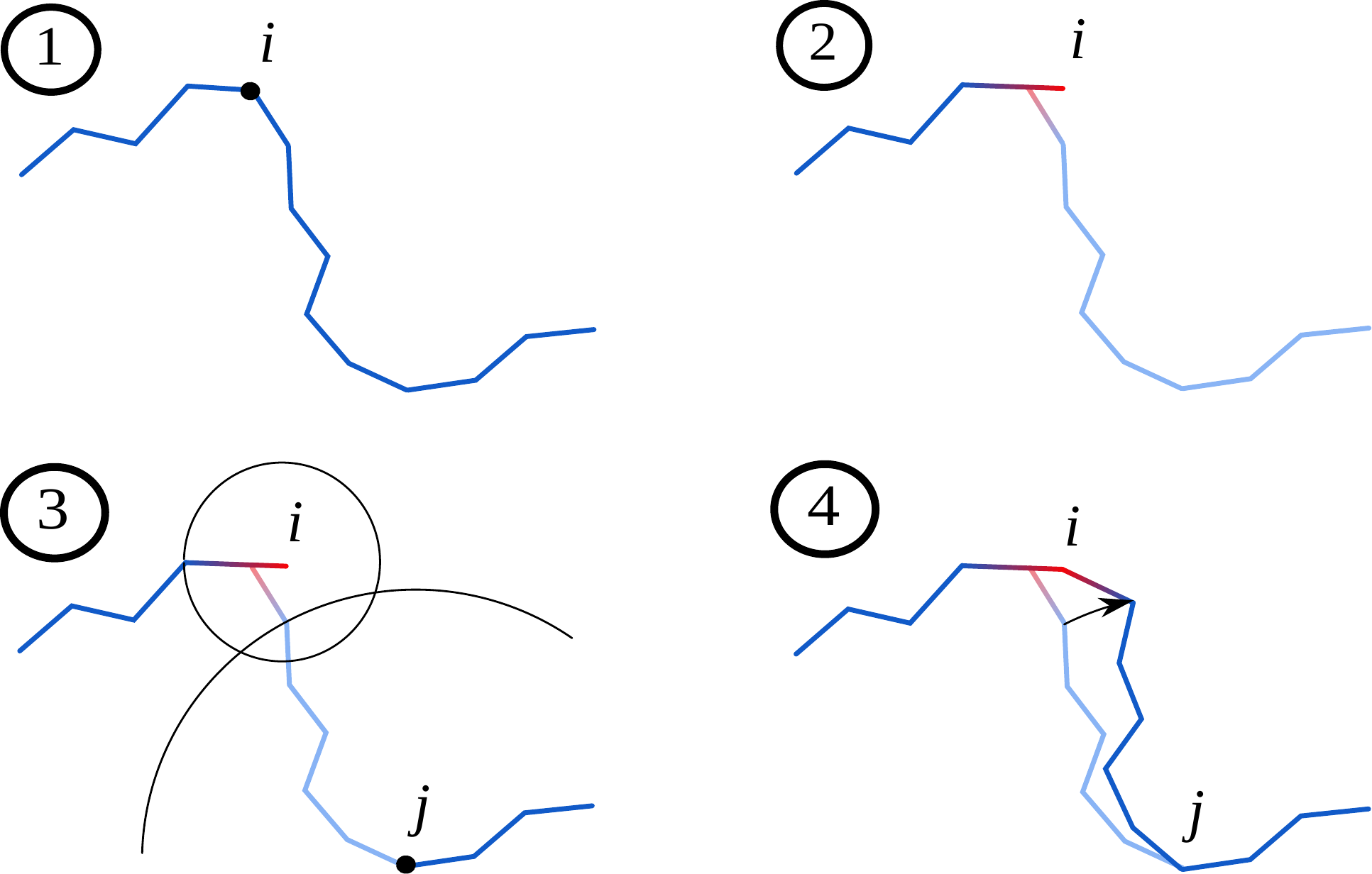}
    \caption{Structural transition: pick a random site $i$ (1).
    In this case, the transition is a denaturation (red), so the length increases (2).
    Pick another site $j$ at random and find a suitable rotation for the block between $i$ and $j$ (3).
    Perform the rotation of the block and change the length of the cylinder on the right of $i$ (4).}
    \label{fig_transition}
	\end{center}
\end{figure}

First, pick a site $i$ at random, whose state is going to be changed.
Decide with probability $1/2$ whether the first length adjustment will be performed towards decreasing indices or towards increasing indices; here, we consider for example the case of increasing indices (see Figure~\ref{fig_transition}). Second, stretch or shrink the cylinder $i-1$ to its new length $l_{i-1}=n(a_{i-1}+a_i)/2$.
Now, the cylinder $i$ has to be displaced in order to match the new position of the site $i$, and its length also has to be changed to $l_i=n(a_i+a_{i+1})/2$.
To this end, choose a block with a random size between the sites $i+1$ and $j>i+1$ and find a rotation which, when applied to this block, will bring the site $i+1$ at the suitable distance $l_i$ from the site $i$.
More precisely, find first the intersection points between the three following objects: i) the sphere centered around the new site $i$ with radius $l_i$ (i.e. the possible positions of the site $i+1$ given the length of the cylinder $i$), ii) the sphere centered around the site $j$ with radius the initial distance between $i+1$ and $j$ (i.e. the possible positions of the site $i+1$ after a rotation of the block) and iii) the plane defined by the cylinders $i-1$ and $i$ (to ensure a minimal deformation of the conformation and thus minimize the number of rejections).
If the intersection does not exist (which can happen when the cylinders are shrinked), reject the move. Otherwise, choose from the two possible points (intersections of the black circles in Figure~\ref{fig_transition}) the one which minimizes the variation in the bending angle between cylinders $i-1$ and $i$ (on the right in that example).

In this context, the final MC method is almost identical to that without structural transitions, the only difference being that at each step, one has to choose with some fixed probability the kind of elementary move to be performed (rotation or structural transition) (Figure~\ref{fig:MC}) and to consider the energy provided by Eq.~\ref{eq:Es}. In this regard, depending on the parameters of the problem (the probability of occurrence of denaturation bubbles), it may prove more efficient to bias the probabilities in favor of one kind of move, as long as the resulting distribution of conformations does not depend on this implementation detail.

\section{Notes}

\begin{enumerate}
\item \label{coarsegraining}
{\it Coarse-graining.}
The level of coarse-graining should be as high as possible to speed up the simulations, but it is constrained by several factors. First, the effective radius $r_e$ imposes a minimal length for the cylinders, which is typically equal to $2r_e$. In the absence of bending energy ($\ell_p=0$), cylinders shorter than this distance indeed produce artefacts because all the cylinders closer than $2r_e$ along the chain are ignored when looking for collisions, but cylinders separated by just a bit more than $2r_e$ are not, and require that all the cylinders in between be almost aligned in order to avoid a collision. This leads to an unexpectedly rigid polymer (thus, the higher $\ell_p$ is, the less important this effect).

The required level of discretisation also depends on the exact properties one is interested in.
For example, 5 cylinders per $\ell_p$ has been shown to be sufficiently accurate to study the extension or the torque of a stretched supercoiled molecule~\cite{Vologodskii1997}, but measurements of finer details, such as the number of plectonemes, requires a higher resolution of 10 cylinders per $\ell_p$~\cite{Lepage2015}.
\item \label{global}
{\it Speed-up using a global energy.}
The simulations can also be sped up by using a thermodynamically equivalent model featuring a global torsional energy~\cite{Vologodskii1997}, all the local fluctuations of the twist being integrated out~\cite{Gebe1995}:
\beq
E(\mathcal{C})=\frac{k_BT}{2}\sum_{i=1}^N\frac{\ell_p}{na}\theta_i^2+2\pi^2k_BT\frac{C}{aN}(\Tw-\Tw_0)^2-fz
\eeq
where $\Tw_0$ is the total twist at rest for the molecule.
\item \label{lk}
{\it Linking number.}
At any time, it is helpful (if not necessary) to check \emph{a posteriori} that the final conformation of the chain has the expected linking number and is unknotted (Note~\ref{knots}).
The conservation of $\Lk$ is verified by computing the sum $\Tw+\Wr$, where the writhe $\Wr$ is given by~\cite{Klenin2000}:
\begin{equation}
\Wr=\frac{1}{2\pi}\sum_{i=2}^N\sum_{j<i}\Omega_{ij}
\end{equation}
To obtain $\Omega_{ij}$, define:
\begin{gather}
a_0=\frac{1}{\sin^2\beta}(\vec{r}_j-\vec{r}_i)(\vec{t}_i\times\vec{t}_j)\\
a_1=\frac{1}{\sin^2\beta}(\vec{r}_j-\vec{r}_i)(\cos\beta\vec{t}_j-\vec{t}_i)\\
a_2=\frac{1}{\sin^2\beta}(\vec{r}_j-\vec{r}_i)(\vec{t}_j-\vec{t}_i\cos\beta)\\
F(x,y)=-\arctan\frac{xy+a_0^2\cos\beta}{a_0(x^2+y^2-2xy\cos\beta+a_0^2\sin^2\beta)^\frac{1}{2}}
\end{gather}
where $\beta$ is the angle between $\vec{t}_i$ and $\vec{t}_j$.
Then:
\begin{equation}
\Omega_{ij}=F(a_1+l_i,a_2+l_j)-F(a_1+l_i,a_2)-F(a_1,a_2+l_j)+F(a_1,a_2)
\end{equation}
\item \label{IC}
{\it Initial conditions.}
For circular molecules, the initial conformation may be as simple as a planar, $N$-sided regular polygon (approximating a circle), since the rotations will quickly desorganize this structure. For linear chains, a straight line is however a bad idea: the axis of rotation of any inner block will pass exactly through the cylinders, and thus will have no effect.
A solenoid is a better starting conformation, provided the pitch is high enough so that the cylinders have room to move without too many collisions occurring. Compute the writhe of this conformation (see Note~\ref{lk}) in order to know the initial twist.
\item \label{acceptance}
{\it MC parameters and acceptance rates.}
The maximal block size ($M$) and the maximal angle ($\alpha$) for the rotations should be chosen such that the acceptance ratio during the simulation is neither close to 1, nor to 0. For a ratio close to 1, energies and conformations almost never vary; larger blocks and larger angles are then more efficient to generate uncorrelated conformations. For a ratio close to 0,  most computation time is wasted into moves that are almost always rejected; smaller blocks and angles would then produce less collisions and crossings but also lower energy variations. Note also  that the larger the surface swept by the block is, the longer the crossing-detection routine. Typically, in absence of chain confinement, a good compromise consists of blocks that span several persistence lengths (e.g. $Mna\approx20\ell_p$), while $\alpha\approx30^\circ$.
\item \label{ptrs}
{\it Pointers and better performances.}
For better performances, avoid unnecessary copies of the data. In particular, it is not efficient to save a copy of the original chain before attempting a move and restore the copy if the move was rejected. Instead, create only two copies of each cylinder at the beginning of the simulation, and work only with pointers or references to theses copies. One pointer designates the current state of the cylinder while the other one refers to a ``draft''. If the elementary move is accepted, swap both pointers so that the draft becomes the new reference for this cylinder. If the move is rejected, the reference stays unchanged.
\item \label{walls}
{\it Walls for linear molecules.}
During the simulation of a linear molecule, the walls may be crossed in different ways.
First, a rotation may bring a cylinder beyond a wall, such that it is necessary to check that any moving site ends its rotation in between the walls.
Second, the trespassing of a cylinder may happen during a rotation, even though both its initial and final positions are valid.
To prevent this scenario, ensure that the extremal coordinates (along the axis of the stretching force) of any site during a rotation remain within the walls.
Third, the alternative type of rotation used with linear molecules allows the walls to move. If one of this wall movement goes toward the center of the chain, check that the non-moving cylinders remain between the walls.
\item \label{cross_sense}
{\it Crossing detection.}
Due to the rounding errors inherently associated with floating-point computation, the detection algorithm may exceptionally fail to report a crossing, which results in an instantaneous variation of the linking number by $\pm2$, and the possibility that a knot is formed.
In order to mitigate that risk, count every near-miss as a crossing (reject the move every time a cylinder is found too close (within some arbitrary threshold) to the surface swept by a moving cylinder).
The few false-positive crossings will not affect the final results of the simulations.
Other methods can be used to mitigate that risk, such as using higher-precision floating-point numbers or reducing the amplitude of the rotations, but all of them will come at a cost in performance.
\item \label{knots}
{\it Knots.}
No absolute invariant is known to discriminate between any two different knots, however a few partial invariants have proven to be useful in practice.
In particular, the Alexander polynomial~\cite{Alexander1928} has been used in this type of simulations because knots sharing the same Alexander polynomial as unknotted conformations are complex enough to be unlikely to appear during simulations. Moreover, the complete polynomial does not need to be computed: its value at $x=-1$ is sufficient to discriminate unknotted conformations from the simplest knots~\cite{Vologodskii1974}.
\end{enumerate}

\section*{Remarks}

The final published version will be available on Springerlink.com

\bibliographystyle{ieeetr}
\bibliography{biblio}

\end{document}